# Observation of perovskite topological valley exciton-polaritons at room temperature

Feng Jin[1,#], Subhaskar Mandal[1,#], Zhenhan Zhang[1], Jinqi Wu[1], Wen Wen[1], Jiahao Ren[1], Baile Zhang[1,2], Timothy C.H. Liew[1, *], Qihua Xiong[3,4,5,6, *], and Rui Su[1,7,*]

[1]Division of Physics and Applied Physics, School of Physical and Mathematical Sciences, Nanyang Technological University, 21 Nanyang Link, Singapore 637371, Singapore

[2]Centre for Disruptive Photonic Technologies, Nanyang Technological University, 21 Nanyang Link, Singapore 637371, Singapore

[3]State Key Laboratory of Low-Dimensional Quantum Physics and Department of Physics, Tsinghua University, Beijing 100084, P.R. China

[4]Frontier Science Center for Quantum Information, Beijing, 100084, P. R. China.

[5]Collaborative Innovation Center of Quantum Matter, Beijing, 100084, P.R. China.

[6]Beijing Academy of Quantum Information Sciences, Beijing 100193, P. R. China.

[7]School of Electrical and Electronic Engineering, Nanyang Technological University, 639798, Singapore

[#]These authors contributed equally to this work.

[*]E-mails: : TimothyLiew@ntu.edu.sg (T.C.H.L.); Qihua_xiong@tsinghua.edu.cn (Q. X.); surui@ntu.edu.sg (R. S.).

**Abstract**

Topological exciton-polaritons are a burgeoning class of topological photonic systems distinguished by their hybrid nature as part-light, part-matter quasiparticles. Their further control over novel valley degree of freedom (DOF) has offered considerable potential for developing active topological optical devices towards information processing. However, the experimental demonstration of propagating topological exciton-polaritons with valley DOF remains elusive at room temperature. Here, employing a two-dimensional (2D) valley-Hall perovskite lattice, we report the experimental observation of valley-polarized topological exciton-polaritons and their valley-dependent propagations at room temperature. The 2D valley-Hall perovskite lattice consists of two mutually inverted honeycomb lattices with broken inversion symmetry. By measuring their band structure with angle-resolved photoluminescence spectra, we experimentally verify the existence of valley-polarized polaritonic topological kink states with a large gap opening of ~ 9 meV in the bearded interface at room temperature. Moreover, these valley-polarized states exhibit counter-propagating behaviors under a resonant excitation at room temperature. Our results not only expand the landscape of realizing topological exciton-polaritons, but also pave the way for the development of topological valleytronic devices employing exciton-polaritons with valley DOF at room temperature.

**Introduction**

Microcavity exciton-polaritons are bosonic quasiparticles resulting from the superposition of semiconductor excitons and microcavity photons in the strong coupling regime[1,2]. Being part-light, part-matter, they fuse the advantages from their constituent components, such as a low effective mass, strong optical nonlinearity, etc[3,4]. Leveraging these unique properties, exciton-polaritons have emerged as fascinating platforms, not only for investigating collective quantum phenomena at elevated temperatures, for instance, polariton Bose−Einstein condensation[5], superfluidity[6], quantum vortices[7], but also for realizing high-performance optoelectronic devices, including low-threshold polariton lasers[5,8-11], all-optical polaritonic switches[12-14] and transistors[15,16]. Recent advances in trapping potentials have further unlocked avenues to precisely manipulate exciton-polaritons with various intriguing Hamiltonians[17-22]. In particular, inspired by the integer quantum Hall effect in condensed matter, the extension of topological concepts has largely sparked the emerging field of topological exciton-polaritons[23-31], endowing novel functionality of topological protection to exciton-polariton systems. Their hallmark feature is the appearance of non-trivial topological polariton edge states with immunity against perturbations, which not only provides exciting possibilities for investigating the interplay between bosonic phenomena and topology, but also promises polaritonic devices a more robust future.

Early investigations into topological exciton-polaritons date back to the theoretical predictions based on various potential landscapes with or without a strong magnetic field, giving rise to localized or propagating edge states[32-35]. Based on Su-Schrieffer-Heeger polariton lattices, localized topological exciton-polaritons was experimentally achieved with GaAs at cryogenic temperatures[23], and further pushed into novel semiconductors hosting more robust polaritons at room temperature[36], such as lead halide perovskites and organics[26,28]. In addition to localized states, the pursuit of propagating topological exciton-polaritons with novel DOF is highly desirable, as they hold significant promise as information carriers capable of transmitting signals robustly against backscattering towards scalability and reliability. By either breaking or preserving the time-reversal (TR) symmetry, propagating edge states were only experimentally achieved in 2D GaAs honeycomb lattices in the quantum Hall phase[24] and a monolayer $WS_2$ coupled to an analogous 2D quantum spin Hall photonic crystal[25,27], respectively. However, constrained by the stringent conditions of forming polaritons, earlier demonstrations in these systems can only work at cryogenic temperatures. To the best of our knowledge, the realization of propagating topological exciton-polaritons working at room temperature remain elusive but highly crucial towards potential applications, which demands platforms with more robust polaritons.

In addition to quantum Hall and quantum spin Hall systems, the quantum valley Hall effect appears as another promising mechanism to support propagating topological states, named valley kink states, which have witnessed substantial successes in various research fields, including electrons in condensed matter systems[37,38], electromagnetic waves in photonic systems[39-44] and acoustic/elastic waves in phononic systems[45-50]. By breaking the inversion symmetry, the valley degeneracy is lifted and a pair of counter-propagating valley kink states with opposite valley-polarization could arise at the non-trivial interface between two domains with opposite valley-Chern numbers[51,52], in the absence of inter-valley scattering.

Importantly, they emerge without the need to break the TR symmetry and provide an extra degree of freedom (DOF) based on valleys. This additional DOF endows topological states with distinct valley polarizations and transport behaviors, holding great promise to store and carry information towards valleytronic applications[53-58]. However, such phenomenon has not been experimentally achieved in any exciton-polariton systems yet.

In this study, by taking advantage of CsPbBr$_3$ perovskite microcavities hosting robust polaritons at room temperature, we report the experimental realization of valley-polarized topological polariton kink states and valley-dependent propagation in a 2D valley-Hall perovskite lattice working at room temperature. By designing honeycomb lattices with broken inversion symmetry, we experimentally construct a 2D valley-Hall perovskite lattice with a bearded interface between two mutually inverted lattices. Through mapping the momentum-space and real-space photoluminescence spectra, we experimentally demonstrate compelling evidence to validate the emergence of the valley-polarized polaritonic topological kink states in both momentum space and real space. Furthermore, under a resonant pulsed excitation, we take advantage of the valley DOF and observe valley-dependent propagation of the topological kink states, where some polaritons can propagate a macroscopic distance of 8.2 μm at room temperature.

**Results**

**Mechanism of the topological 2D valley-Hall perovskite lattice**

The inherent symmetries of honeycomb lattices result in valley degeneracy, introducing an additional DOF akin to spins and paving the way for the burgeoning field of valleytronics. Motivated by this, we begin by theoretically examining a system of exciton-polariton micropillars arranged in a honeycomb lattice (lattice periodicity, $a$ = 0.85 μm; micropillar diameter, $d$ = 0.6 μm). The band structure of such system along the high symmetry line ($\Gamma \to M \to K \to \Gamma$) of the Brillouin Zone (BZ) is illustrated by the red dashed lines in Figure 1a. Analogous to electronic graphene, this system also exhibits gapless Dirac points at the two valleys, which are protected by both TR and inversion symmetries and remain gapless unless either of these symmetries is disrupted. While TR symmetry breaking has been explored in polariton systems[24], our focus here is on breaking inversion symmetry by selecting different pillar diameters corresponding to the two sublattices (small micropillar diameter, $d_1$ = 0.5 μm; big micropillar diameter, $d_2$ = 0.7 μm). The brown solid lines in Figure 1a showcase the band structure of the inversion symmetry broken system along the high symmetry line of the BZ, revealing the gapping of the Dirac point and the emergence of a bulk bandgap of approximately 5 meV.

To achieve topologically protected valley kink states, we examine a structure comprising two domains of the inversion symmetry broken system, with one domain obtained by reflecting the other across the $x = 0$ axis, as depicted in Figure 1b. While the entire structure maintains periodicity along the $y$-direction, periodicity is disrupted along the $x$-direction at $x = 0$, forming a bearded interface between the two domains. Despite having identical band structures, as shown in Figure 1a, the domains exhibit distinct topological properties. To recognize these properties, we compute the Berry curvature $\mathcal{F}(k_x, k_y)$

for the two domains[59]

$$\mathcal{F}(k_x, k_y) = \nabla_k \times \mathcal{A}(k_x, k_y),$$

with $\mathcal{A}(k_x, k_y) = <u(k_x, k_y)|i\nabla_k|u(k_x, k_y)>$ denoting the Berry connection comprising the Bloch modes $|u(k_x, k_y)>$. Figure 1c illustrates the Berry curvature for the two domains, indicating non-zero values near the valleys with opposite characteristics for each domain. Additionally, a valley-Chern number can be defined as

$$C = \frac{1}{2\pi} \iint \mathcal{F}(k_x, k_y) \, dk_x dk_y,$$

with the integral taken over half of the BZ. For domain 1, $C_K^1 = -C_{K'}^1 = 1/2$, while for domain 2, $C_K^2 = -C_{K'}^2 = -1/2$, resulting in a difference $\Delta C_K = +1$ and $\Delta C_{K'} = -1$ between the two domains. According to the bulk-boundary correspondence, the interface between the domains will host counter-propagating topologically protected states at the two valleys. To verify this, we compute the projected band structure of the system in a strip geometry, where the system is periodic along the $y$-direction. In Figure 1d (left), the strip band structure is depicted, revealing topologically protected counter-propagating valley-polarized states inside the bulk bandgap. Additionally, the spatial profile of one of the topological modes, shown on the right panel of Figure 1d, further confirms their appearance at the interface.

**Experimental characterization of bulk and topological valley kink states in the valley-Hall lattice**

Experimentally, we fabricate the valley-Hall lattice by etching the spacer layer of $CsPbBr_3$ perovskite microcavity, in which the interface is well-aligned along the crystal axes of the perovskite, as illustrated in Figure 2a (Refer to methods for more details). Figure 2b presents the typical scanning electron microscopy (SEM) image of the valley-Hall lattice, which shows the bearded interface with two domains. To demonstrate the emergence of the topological valley kink states in the polariton valley-Hall lattice, we perform the momentum-space and real-space photoluminescence characterizations with a non-resonant continuous-wave laser excitation of 2.713 eV at room temperature (Methods). We first probe the topological kink states in a valley-Hall lattice sample with a detuning of ~ -115 meV and similar results can also be observed in other samples with different detunings (Supplementary Section 2). Near the Dirac points ($E= \sim 2.296$ eV), we examine the behavior in the momentum space (Figure 2c) and it exhibits a clear hexagonal shape, which corresponds to the BZ with typical $K/K'$ valleys. Furthermore, we characterize the band structure of the valley-Hall lattice from the bulk area and the bearded interface area, framed by the white and red dashed lines in Figure 2b, respectively. Figure 2d illustrates the energy-wavevector dispersion from the bulk area along the $k_y$ direction at $k_x = 0$ μm$^{-1}$ (along $K \to \Gamma \to K'$). With the strong confinement in the polariton pillars, the $s$-mode polaritons couple together to form the $s$ energy band, ranging from 2.262 eV to 2.320 eV, which constitutes the main region of the band structure. Within the $s$ band, we experimentally observe a gap opening of ~5 meV at two nonequivalent $K$ and $K'$ points (framed by black dashed lines), as a result of the broken inversion symmetry in the valley-Hall lattice. Subsequently, we move to the bearded interface and characterize its dispersion along $K \to \Gamma \to K'$ (white dashed line in Figure 2c). As illustrated in Figure 2e, in addition to the gap opening of ~ 9 meV, a pair of topological kink states emerge inside the gap (indicated by the red arrow), which agrees well with our theoretical prediction (Figure 1d and Supplementary Section 1). Inside the first BZ (highlighted by the blue dashed lines), the dispersion slopes of the two topological kink states are opposite at different K and K′ valleys, leading to valley-dependent group velocities with opposite propagating directions. In other words, the topological kink states are valley-polarized and locked to one propagating direction in the absence of inter-valley scattering. The emergence of the topological kink states can be further clarified by detecting the real space profiles at the bulk state ($E=$ 2.285 eV) and the topological state ($E=$ 2.296 eV), indicated by blue and red arrows in Fig. 2e, respectively. Notably, the emission from the bulk state is mainly from the bulk valley domains (left panel in Figure 2f), whereas the emission from the topological kink state appears mainly at the domain wall of the valley-Hall lattice (right panel in Figure 2f).

To enhance the understanding of the valley-polarized topological kink states, we also collect the angle-resolved photoluminescence spectra along $K' \rightarrow M \rightarrow K$ (yellow dashed line in Figure 2c) by translating the Fourier lens. Figure 3a illustrates the dispersion from the bulk area along $k_y$ at $k_x = 3.2$ μm$^{-1}$, revealing a similar bandgap opening of ~ 5 meV at ~ 2.296 eV without any states inside (highlighted by the black dashed lines). As a clear comparison, the band structure from the bearded interface exhibits a pair of additional dispersions with opposite slopes inside the gap, which correspond to the valley-polarized topological kink states (indicated by red arrow). The existence of the topological kink states can also be experimentally confirmed through comparing the energy-resolved spatial images collected along the bulk domain (top, Figure 3c) and the domain wall (bottom, Figure 3c), as indicated by the green and orange dashed lines in Figure 2b, respectively. Specifically, only at the domain wall of the valley-Hall lattice, the topological kink states can be found inside the topological gap at 2.296eV, highlighted by red dashed lines in Figure 3c. Our results collectively validate the emergence of valley-polarized topological kink states localized at the domain wall in the polariton valley-Hall lattice.

**Observation of valley-dependent propagation with topological exciton-polaritons**

One of the unique advantages in valley-Hall polariton lattices is the valley DOF, with which the valley-polarized kink states are locked to one specific propagating direction in the absence of inter-valley scattering. We further experimentally demonstrate such valley polarization and valley-dependent topological polariton propagation by selectively and resonantly exciting the valleys in a transmission configuration at room temperature. As illustrated in Figure 4a, a lattice sample with a detuning of ~ -190 meV is resonantly excited with a pulsed laser from the back of the microcavity and we collect both the momentum space and real space spectra from its front side. In order to find out the exact valley positions, we also collect the photoluminescence dispersion from the bearded interface of the valley-Hall lattice along $K \rightarrow \Gamma \rightarrow K'$, revealing that the topological kink states locate at $E = $ ~ 2.231 eV with $k_y = \pm 3.8$ μm$^{-1}$ and $k_x = 0$ μm$^{-1}$, as illustrated in Figure 4b (left) and 4e (right). Subsequently, we tune the pumping laser beam to resonantly excite the $K$ ($K'$) valley-polarized topological polaritons in the domain wall, respectively. As shown in Figure 4b (right), when one $K$ valley-polarized topological kink state at $k_y = +3.8$ μm$^{-1}$ and $k_x = 0$ μm$^{-1}$ is excited, we observe scattering among the same valleys and negligible intervalley scattering, where the 2D momentum-space spectra exhibits bright spots only at equivalent $K$ valleys but no emission at the $K'$ valleys (Figure 4c). In the meantime, as shown in Figure 4d (left), these $K$ valley-polarized topological polaritons exhibit long-range propagation along the domain wall in the $+y$ direction, where some polaritons can propagate as far as 8.2 μm at room temperature.

In sharp contrast, we observe distinct scenarios in the $K'$ valleys. As shown in Figure 4e (left), we resonantly excite the $K'$ valley at $k_y = -3.8$ μm$^{-1}$ and $k_x = 0$ μm$^{-1}$, which corresponds to the $K'$ valley-polarized topological polariton states. In the 2D momentum space, the other equivalent $K'$ valleys in the first BZ also light up and no emission can be observed from the $K$ valleys, suggesting the absence of intervalley scattering. Furthermore, the trajectory of the $K'$ valley-polarized topological polaritons exhibit similar propagation behaviors along the domain wall, but with an opposite direction of -$y$. Similar evidence can also be observed from the energy-resolved spatial images when $K$ ($K'$) valley-polarized topological polaritons are excited respectively (Supplementary Section 3). Our observations strongly suggest the existence of valley polarization and valley-dependent propagation in our polariton valley-Hall lattices at room temperature.

**Discussion**

In summary, we have demonstrated the realization of valley-polarized topological exciton-polaritons and their valley-dependent propagation in 2D valley-Hall perovskite lattices at room temperature. By designing a bearded interface between two mutually inverted polariton honeycomb lattices with broken inversion symmetry, we theoretically predict and experimentally verify the existence of topological exciton-polariton kink states. Our experimental results collectively demonstrate that a pair of valley-polarized topological kink states are spectrally locked to $K$ ($K'$) valleys inside the topological gap and spatially confined at the bearded interface of the valley-Hall lattice. Additionally, under a resonant excitation, we further showcase their valley polarization and valley-dependent propagation of the topological kink state polaritons along the domain wall of the valley-Hall lattice, where some polaritons can propagate as far as 8.2 μm at room temperature. Our work introduces the valley DOF into the topological polariton system, paving the way for designing novel valleytronics polariton devices for communication and computing in ambient conditions. With the unique advantages of exciton-polaritons, we also anticipate exciting opportunities for designing low-threshold valley-addressable topological lasers and nonlinear switchable topological devices.

## Methods

**Perovskite valley-Hall lattice fabrication**

The perovskite microcavity consists of a bottom DBR, an all-inorganic perovskite (CsPbBr$_3$) nanoplatelet, a ZEP lattice pattern and a top DBR. In detail, 15.5 pairs of TiO$_2$/SiO$_2$ layers are deposited on the silicon or silica wafer as bottom DBR substrate for reflection and transmission configurations, respectively, which are fabricated by an electron beam evaporator (Cello 50D). The 65 nm-thick single-crystalline CsPbBr$_3$ is synthesized via a chemical vapor deposition method as described before[60], and then transferred onto the bottom DBR substrates through the dry-transfer process with Scotch tape[61]. Next, 85 nm-thick ZEP520A layers are spin-coated on the samples before depositing the top DBRs, and then patterned into a 2D valley-Hall lattice by standard electron beam lithography. Lastly, 8.5 pairs of TiO$_2$/SiO$_2$ layers are deposited as top DBRs by the electron beam evaporator.

**Optical spectroscopy characterizations**

The energy-resolved momentum-space and real-space photoluminescence are detected by a home-built angle-resolved spectroscopy setup with Fourier optics. The emission from the microcavity is collected by a 50× objective lens (NA = 0.75) and sent to a 550 mm focal length spectrometer (Horiba iHR550) with a grating (600 lines/mm) and a liquid nitrogen cooled CCD (256 × 1,024 pixels). The perovskite microcavity is non-resonantly pumped by a continuous wave laser (457 nm) with a pump spot of ~25 μm. For transmission configuration, the microcavity is resonantly pumped by a pulsed laser (556 nm wavelength center, 1 kHz repetition rate and 100 fs pulse duration), and the pumping beam is spectrally filtered by using a band-pass filter. In addition, a slit and an angle-variable lens are employed to adjust the incidence angle of the pulsed pumping source.

**Theoretical calculations**

To model the system, we utilize coupled Schrödinger equations within the mean-field approximation. These are formulated as follows:

$$i\hbar \frac{\partial \phi_{ph}(r,t)}{\partial t} = \left[-\frac{\hbar^2}{2m_{ph}}\left(\frac{\partial^2}{\partial x^2} + \frac{\partial^2}{\partial y^2}\right) + U(r)\right]\phi_{ph}(r,t) + \frac{g_0}{2}\phi_{ex}(r,t),$$

$$i\hbar \frac{\partial \phi_{ex}(r,t)}{\partial t} = E_{ex}\phi_{ex}(r,t) + \frac{g_0}{2}\phi_{ph}(r,t).$$

Here, $\phi_{ph}$ and $\phi_{ex}$ represent the mean-field wave functions of photons and excitons, respectively. $m_{ph} = 1.7 \times 10^{-5} m_e$ denotes the isotropic cavity photon mass, with $m_e$ representing the free electron mass. $U(r)$ represents the in-plane photon confinement potential, as illustrated in Figure 1b, while $g_0$ indicates the Rabi splitting. Due to the heavy exciton mass, excitons are considered dispersion-less, with energy $E_{ex}$. To derive the band structures in Figure 1d, we apply the Bloch theorem and the plane wave expansion method. Additional parameters include $g_0$=120 meV and $E_{ex}$=2.407 eV.

**Acknowledgements**

R.S. and T. C.H. L. gratefully acknowledge funding support from the Singapore Ministry of Education via the AcRF Tier 2 grant (MOE-T2EP50222-0008) and Tier 1 grant (RG80/23). Q.X. gratefully acknowledges strong funding support from National Natural Science Foundation of China (grant No. 12020101003 and 92250301). R.S. also gratefully acknowledges funding support from Nanyang Technological University via a Nanyang Assistant Professorship start-up grant. R.S. and B.L. Zhang gratefully acknowledge funding support from the Singapore National Research Foundation via a Competitive Research Program (grant no. NRF-CRP23-2019-2670007).

**Author contributions**

R.S., Q.X. and T.C.H.L. supervised the project. F.J. and J.H.R. synthesized the perovskite materials. F.J. fabricated the devices with the help of J.Q.W. and Z.H.Z. F.J. performed all the optical spectroscopy measurements. T.C.H.L. and S.M. conceived the lattice model with input from R.S. and Q.X. S.M. performed the theoretical calculations. R.S., F.J. and S.M. wrote the manuscript with the inputs from all the authors.

**Competing interests**

The authors declare that they don't have competing interests.

**Data availability**

All experimental data that support the plots within this paper are available from the corresponding author upon reasonable request.

**Code availability**

The codes are available from the corresponding author upon reasonable request.

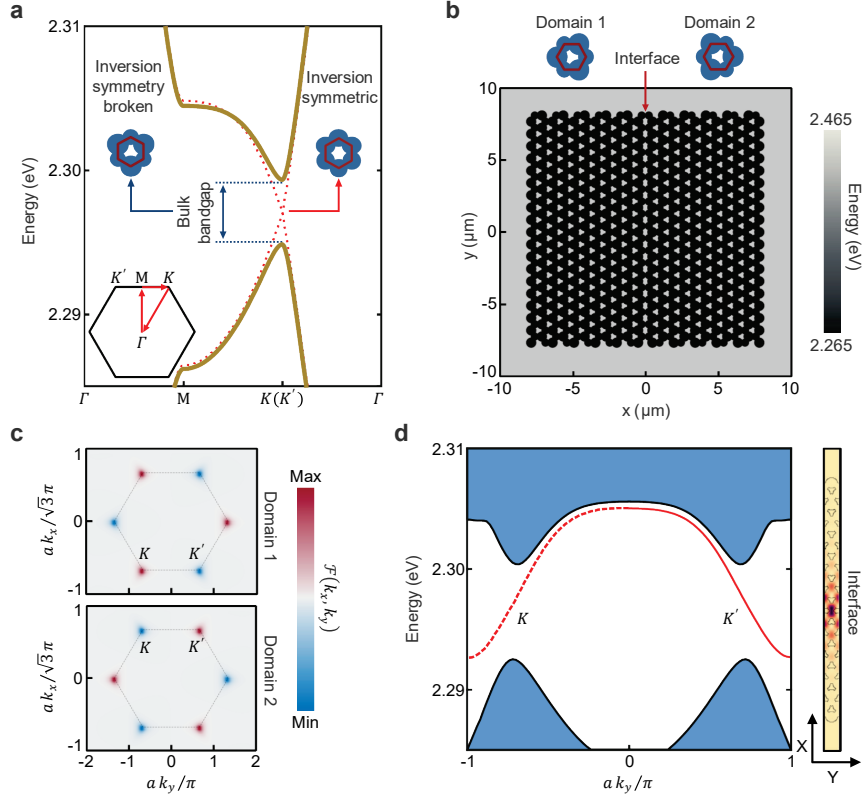

**Fig. 1| Mechanism of the topological valley-Hall polariton lattice. a,** Band structure of exciton-polariton micropillars arranged in a honeycomb lattice. The dashed and solid lines correspond to the inversion symmetry-preserved and inversion symmetry-broken system, respectively. The inset at the bottom shows the high-symmetry line of the BZ along which the band structures are calculated. **b,** Two domains of the honeycomb lattice with broken inversion symmetry. One domain is a reflected copy of the other to create an interface at $x = 0$. The unit cells are shown above the lattice. **c,** Numerically calculated Berry curvature of the two domains, indicating opposite Berry curvature in the two domains, resulting in a difference of $\Delta C_{K(K')} = \pm 1$ in valley-Chern numbers. **d,** Projected band structure of the system in **b**, with the $y$ direction taken as periodic. Inside the bulk bandgap, counter-propagating valley-polarized modes located at the interface appear, which are protected by the non-trivial valley-Chern number. The spatial profile of one of the interface modes is shown on the right. $a = 0.85$ μm is the periodicity along the $y$ direction.

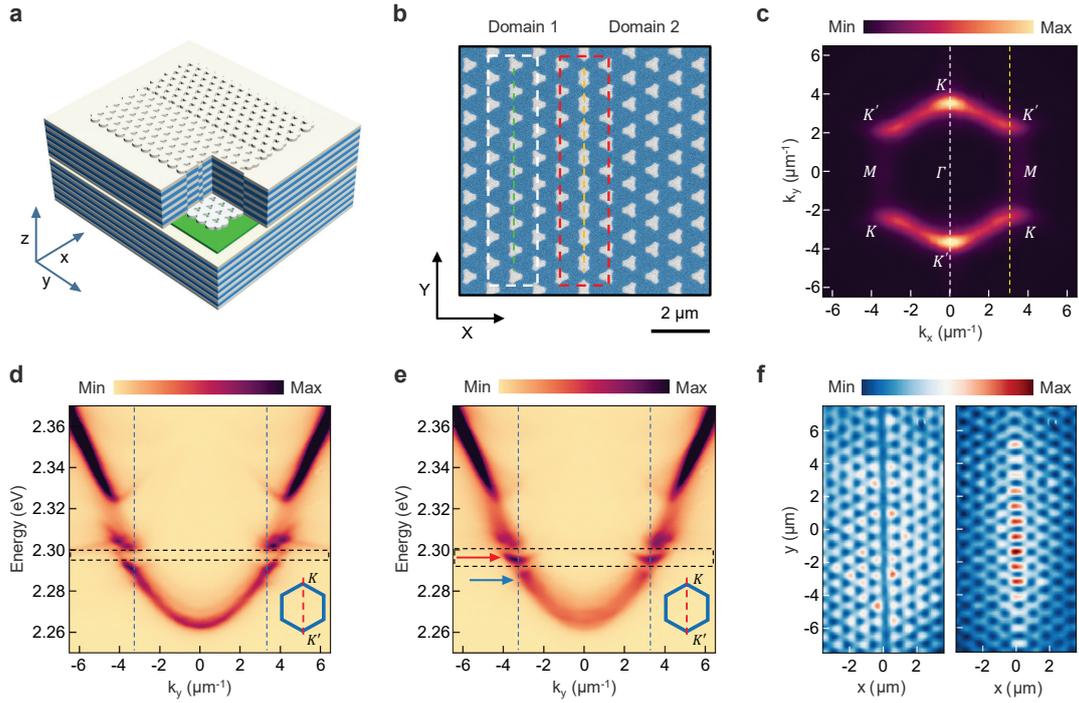

**Fig. 2| Observation of topological valley exciton-polaritons at room temperature. a,** Schematic representation of the perovskite topological valley-Hall lattice microcavity structure, where the lattice is created by patterning the ZEP layer and the lattice interface is well-aligned along the crystal axes of the perovskite. **b,** Scanning electron microscopy image of the 2D valley-Hall lattice on the perovskite layer before depositing the top DBR. The white and red dashed lines frame the emission collection areas in **d** and **e**, respectively. The green and orange dashed lines represent the emission collection areas for the energy-resolved spatial images in **Fig. 3c**. **c,** 2D momentum-space emission of the valley-Hall lattice near the Dirac points ($E = \sim 2.296$ eV). The white and yellow dashed lines are at $k_x = 0$ μm$^{-1}$ and $k_x = 3.2$ μm$^{-1}$, respectively. **d,** Momentum-space polariton energy dispersion of the bulk area at $k_x = 0$ μm$^{-1}$ along $K \rightarrow \Gamma \rightarrow K'$. The blue dashed lines represent the first BZ, and the black dashed lines highlight the bandgap opening inside the s band. **e,** Momentum-space polariton energy dispersion of the topological bearded interface area at $k_x = 0$ μm$^{-1}$ along $K \rightarrow \Gamma \rightarrow K'$. The blue dashed lines represent the first BZ. The blue and red arrows indicate the bulk state ($E = 2.285$ eV) and the topological valley kink state ($E = 2.296$ eV), respectively. **f,** Real-space images of the perovskite valley-Hall lattice at the energies of 2.285 eV (left, bulk state) and 2.296 eV (right, topological valley state).

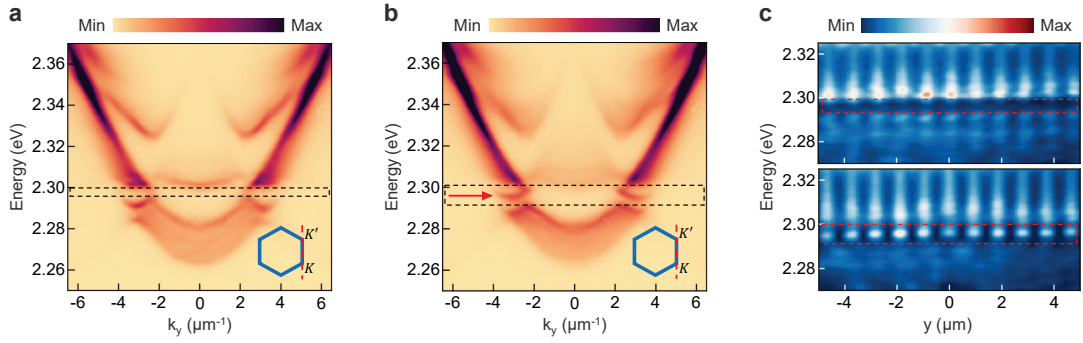

**Fig. 3| Further characterizations of topological valley exciton polaritons at room temperature. a,** Energy-resolved momentum-space polariton dispersion of the perovskite valley-Hall lattice collected from the bulk area at $k_x = 3.2$ μm$^{-1}$ along $K \to M \to K'$. The black dashed lines highlight the bandgap opening. **b,** Energy-resolved momentum-space polariton dispersion of the perovskite valley-Hall lattice collected from the topological bearded interface area at $k_x = 3.2$ μm$^{-1}$ along $K \to M \to K'$. The red arrow indicates the topological valley kink state. **c,** Energy-resolved spatial images collected along the green dashed line (top) and orange dashed line (bottom) in Fig. 2b, respectively. The red dashed lines highlight the topological gap opening inside the $s$ band. The topological valley kink states only emerge at the domain wall of the valley-Hall lattice.

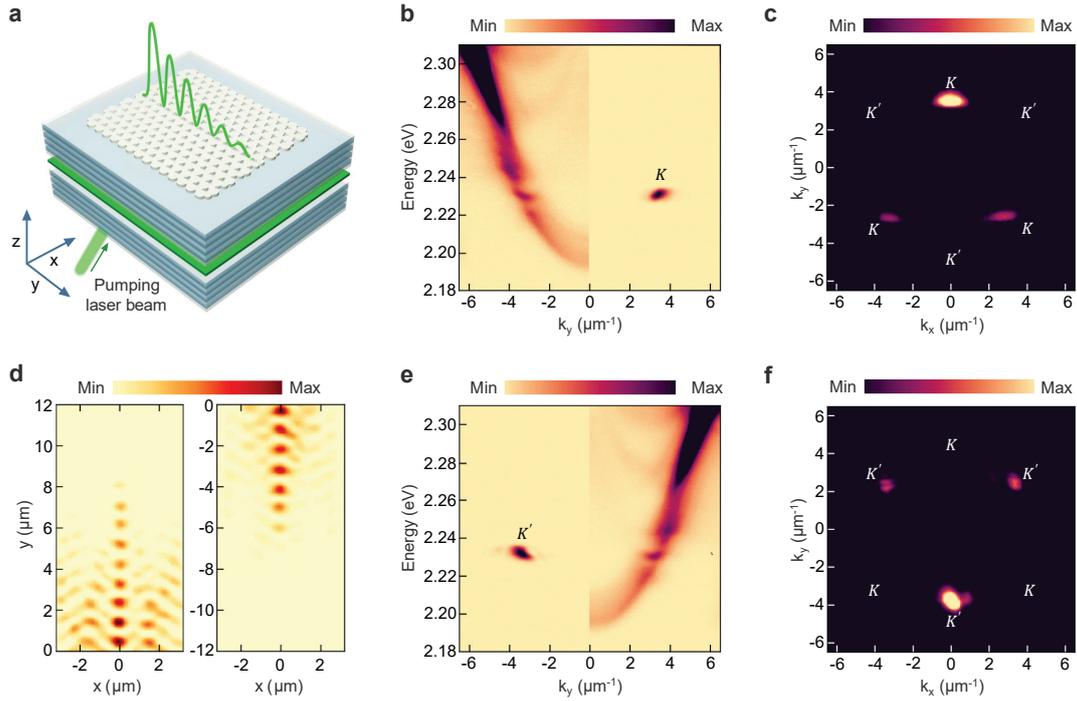

Fig. 4| **Valley-dependent propagation of the topological exciton-polaritons. a,** Schematic of propagating topological valley polaritons in the perovskite valley-Hall lattice. The resonant pumping laser beam is injected with an angle from the back of the microcavity, and the propagation of the topological valley polaritons along the domain wall of the valley-Hall lattice can be observed from the front of the microcavity. **b** (left) and **e** (right)**,** Momentum-space polariton energy dispersions collected from the topological bearded interface area at $k_x = 0$ μm$^{-1}$ with a non-resonant pumping. **b** (right) and **e** (left)**,** Momentum-space polariton energy dispersions of the topological valley polaritons with a resonant excitation at the topological energy state ($E$ = 2.231eV). **c** and **f,** 2D momentum-space photoluminescence spectra of the $K$ ($K'$) valley-polarized topological polaritons under resonant pumping, respectively. **d,** Experimental real-space images of the $K$ ($K'$) valley-polarized topological polaritons propagating along the domain wall of the valley-Hall lattice in $\pm y$ direction, which shows some polaritons can propagate as far as 8.2 μm.